\documentclass[12pt]{iopart}

\expandafter\let\csname equation*\endcsname\relax
\expandafter\let\csname endequation*\endcsname\relax
\usepackage{amsmath,amssymb}
\usepackage{braket}
\usepackage{bm}

\usepackage{graphicx}
\usepackage{epstopdf}

\begin{document}

\title{Squeezed-reservoir decoherence control in strongly coupled optomechanics}

\author{Vaibhav N Prakash$^{1}$ and Aranya B Bhattacherjee$^{2,3}$}

\address{$^{1}$Department of Physics, Mahindra University, Hyderabad-500043, India}
\address{$^{2}$Department of Physics, Birla Institute of Technology and Science, Pilani, Hyderabad Campus, Hyderabad-500078, India}
\address{$^{3}$Department of Physics, School of Applied Sciences, University of Science and Technology, District Ri Bhoi, 793101, Meghalaya, India}

\ead{vaibhav.prakash@mahindrauniversity.edu.in}

\begin{abstract}
Recent advancements in strong single-photon optomechanical coupling also demand a deeper understanding of environmental interactions in this regime. In this regime the standard Lindblad master equation, which is derived in the eigenbasis of the bare optical and mechanical modes, misassigns the dephasing rate. We therefore use the Dressed-State Master Equation (DSME), which is formulated in the eigenbasis of the strongly coupled photon-phonon Hamiltonian, the dressed states of the system. This work investigates the impact of squeezed vacuum and thermal reservoirs on the decoherence of cavity photon Fock states in the strong coupling regime. We demonstrate that decoherence can be effectively controlled by tuning reservoir parameters, with the control mediated through a cavity dephasing term that becomes significant at high temperatures. The findings presented provide critical insights into reservoir engineering for precise control of quantum decoherence, advancing the understanding of strongly coupled optomechanical systems in engineered environments.
\end{abstract}

\vspace{2pc}

\submitto{\PS}

\maketitle

\section{Introduction}\label{sec1}
There is an increasing focus on utilizing optical photons as a foundation for quantum information processing, communication systems, and advanced sensing technologies. Central to these applications are photonic system components, such as single-photon sources \cite{Luc, Ina, Luk, Ama}, generators of nonclassical states of light \cite{Gui, Lin}, interferometers \cite{Alb, Hyo}, and sensors \cite{Tia, Tor, Feng, Ali}, some of which rely on photon-phonon interactions for their functionality. These interactions are particularly critical in the quantum regime, where nonlinear photon-photon interactions mediated by phonons enable phenomena such as photon blockade \cite{Hai, Dong1, kai, Dong2, Rabl, Gao2022}. As optomechanical nonlinearity plays an essential role in the operation of future quantum devices, a deeper understanding of the dynamics of open quantum optomechanical systems in the strong single-photon coupling regime is imperative. Recent theoretical work has shown that even transient driving in this regime can prepare strongly nonclassical mechanical states with Wigner-function negativity \cite{Wise2024}, sharpening the need for accurate descriptions of decoherence.

Previous studies have explored the influence of vacuum and thermal reservoirs on the dynamics of optomechanical systems in the strong single-photon coupling regime \cite{Hu}. These investigations revealed critical insights into how environmental interactions affect system coherence and dynamics. However, the potential of squeezed vacuum and squeezed thermal reservoirs, known to provide additional control over the dynamics of open quantum systems \cite{Hou, Ami, Rog, Chun}, remains underexplored in the strong coupling regime. Squeezed thermal reservoirs introduce new degrees of freedom such as adjustable squeezing parameters and reservoir phase, offering additional flexibility for mitigating decoherence and manipulating quantum states.

In this article, we extend the analysis of \cite{Hu} by treating an optomechanical system in the strong single-photon coupling regime under the influence of squeezed vacuum and squeezed thermal reservoirs \cite{Gon}. We find that for an appropriate choice of squeezing strength and reservoir phase the decoherence rate of cavity photon Fock superpositions can be reduced approximately sevenfold relative to a thermal reservoir at the same temperature. The analysis uses the Dressed-State Master Equation (DSME), which accounts for the modified eigenstates of the strongly coupled system and has been applied in previous studies of such systems \cite{Pon, Cho}. We also propose modifications to the Standard Master Equation (SME) in the interaction picture that incorporate the effects of squeezed thermal reservoirs, providing a complementary framework. The article is organized as follows. Section \ref{sec2} introduces the Hamiltonian and the dressed eigenbasis relevant to the strong photon-phonon coupling regime. Section \ref{sec4} presents the DSME for the reduced density matrix, and Section \ref{sec5} extends it to a squeezed thermal reservoir. The main results are presented and discussed in Section \ref{sec6}. Section \ref{sec7} provides a scaling estimate of the optimal squeezing strength. Section \ref{sec8} concludes.

\section{Model and dressed-state basis}\label{sec2}

\subsection{Non-quadratic Hamiltonian}

\begin{figure}[ht]
\centering
\includegraphics[width=0.9\linewidth]{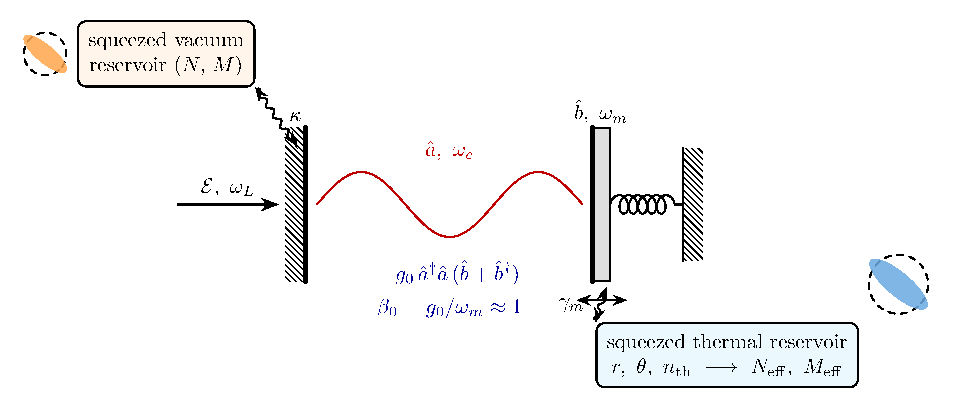}
\caption{Schematic of the system. A Fabry-Perot cavity mode $\hat a$ of frequency $\omega_c$ and linewidth $\kappa$ is driven by a weak laser of amplitude $\mathcal{E}$ and couples through radiation pressure, with single-photon strength $g_0$ in the strong-coupling regime $\beta_0 = g_0/\omega_m \approx 1$, to a mechanically compliant end mirror $\hat b$ of frequency $\omega_m$ and damping rate $\gamma_m$. The cavity input is a squeezed vacuum reservoir with parameters $(N, M)$, and the mechanical mode couples to a squeezed thermal reservoir characterized by the squeezing strength $r$, the phase $\theta$ and the thermal occupancy $n_{\text{th}}$, which enter the master equation through $N_{\text{eff}}$ and $M_{\text{eff}}$. The phase-space icons contrast the phase-insensitive noise of a thermal state (dashed circles) with the squeezed noise ellipses of the engineered reservoirs. The mechanical squeezed reservoir can be realized by the two-tone scheme discussed in Section~\ref{sec5a}.}
\label{fig.schematic}
\end{figure}

This work studies a single-mode optomechanical system in the weak-drive, strong single-photon coupling regime. The strong-drive, weak-$g_0$ regime, in which dressed states arise from a splitting of the normal modes of the non-interacting Hamiltonian \cite{Mar}, is referenced here only for context. We consider a Fabry-Perot cavity with one end mirror that is free to oscillate under the radiation pressure exerted by the cavity photons. Figure~\ref{fig.schematic} sketches the setup together with the two phase-sensitive reservoirs introduced in Section~\ref{sec5}. The mechanical degree of freedom associated with this mirror is modelled as a simple harmonic oscillator, and the optomechanical coupling arises from the radiation pressure force that depends on the photon number in the cavity mode. The Hamiltonian of this system, expressed in a frame rotating at the laser drive frequency $\omega_L$, is
\begin{equation}\label{1}
\hat{H} = -\Delta \, \hat{a}^\dagger \hat{a} + \omega_m \, \hat{b}^\dagger \hat{b} - g_0 \, \hat{a}^\dagger \hat{a} (\hat{b} + \hat{b}^\dagger) + i\mathcal{E} (\hat{a}^\dagger - \hat{a}),
\end{equation}
where $\Delta = \omega_L - \omega_c$ is the detuning between the laser drive frequency $\omega_L$ and the intrinsic cavity frequency $\omega_c$. Here, $\omega_m$ is the natural frequency of the mechanical oscillator. The operators $\hat{a}$ and $\hat{a}^\dagger$ denote the annihilation and creation operators for the cavity mode, while $\hat{b}$ and $\hat{b}^\dagger$ are the annihilation and creation operators for the mechanical mode of the movable mirror. The coefficient $g_0$ is the single-photon optomechanical coupling strength. The final term, $i\mathcal{E} (\hat{a}^\dagger - \hat{a})$, represents the external laser drive of amplitude $\mathcal{E}$. The non-quadratic term $g_0 \, \hat{a}^\dagger \hat{a} (\hat{b} + \hat{b}^\dagger)$ introduces the optomechanical non-linearity essential for phenomena such as photon blockade, mechanical squeezing, and radiation-pressure cooling of the mechanical mode. The drive enters the analysis only through the preparation of the initial cavity state. All results below are obtained in the undriven limit $\mathcal{E} \to 0$, taken in Section~\ref{sec3}, and the initial superposition state that the drive prepares is specified in Equation~(\ref{14}) of Section~\ref{sec6}.

\subsection{Polaron transform and dressed states}\label{sec3}
We now take the undriven limit ($\mathcal{E}\to 0$) in order to identify the eigenstates of the photon-phonon interaction itself. The drive only sets initial conditions for the analysis that follows and does not affect the eigenbasis. Applying a polaron transformation to Equation~(\ref{1}) in this limit gives
\begin{equation}\label{eq:hamiltonian_transformation}
\lim_{\mathcal{E}\to 0}\hat{H}' = \hat{U} \hat{H} \hat{U}^\dagger = \omega_c \, \hat{a}^\dagger \hat{a} + \omega_m \, \hat{b}^\dagger \hat{b} - \frac{g_0^2}{\omega_m} (\hat{a}^\dagger \hat{a})^2,
\end{equation}
where $\hat{U} = e^{-S}$ denotes a unitary operator, with the generator $S$ given by
\begin{equation}
S = (\hat{a}^\dagger \hat{a}) \frac{g_0}{\omega_m} (\hat{b}^\dagger - \hat{b}),
\end{equation}
as shown in \cite{Nun}. Here, $\omega_c$ and $\omega_m$ are the frequencies of the cavity mode and mechanical mode, respectively, while $g_0$ represents the single-photon optomechanical coupling strength. This transformation isolates the intrinsic nonlinear interactions between photons and phonons by removing the linear optomechanical terms.

With the transformed Hamiltonian $\hat{H}'$, we can express the total energy of the system in terms of photon and phonon occupation numbers. The energy of the system is then given by
\begin{equation}
E' = n \omega_c + m \omega_m - \frac{n^2 g_0^2}{\omega_m},
\end{equation}
where $n = \langle \hat{a}^\dagger \hat{a} \rangle$ and $m = \langle \hat{b}^\dagger \hat{b} \rangle$ are the expectation values of the photon and phonon number operators, respectively. The term $-n^2 g_0^2/\omega_m$ captures the nonlinear energy shift of an effective photon-photon interaction. It originates from the photon-phonon coupling $g_0$, which the polaron transformation converts into the Kerr-type term of Equation~(\ref{eq:hamiltonian_transformation}) once the mechanical mode is eliminated. Even in the single-photon regime where $n$ is small, this shift becomes appreciable when the dimensionless coupling $\beta_0 = g_0/\omega_m$ is of order unity, which is the regime of interest here.

The eigenstates of the transformed Hamiltonian $\hat{H}'$ can be represented as product states involving both the photon and phonon components. These eigenstates take the form
\begin{equation}\label{eq:eigenstate_product}
\ket{\Psi} = \ket{n} \otimes \ket{m_{(n)}} = \ket{n} \otimes e^{\hat{N_c} \beta_0 (\hat{b}^\dagger - \hat{b})} \ket{m},
\end{equation}
where $\beta_0 = \frac{g_0}{\omega_m}$ and $\hat{N_c} = \hat{a}^\dagger \hat{a}$ is the photon number operator. Here, $\ket{m_{(n)}}$ represents a displaced mechanical state $\ket{m}$, where the displacement amplitude is proportional to the photon number, given by $n \beta_0$. The term $\ket{n}$ denotes the photon Fock state.

\begin{figure}[ht]
\centering
\includegraphics[width=0.55\linewidth]{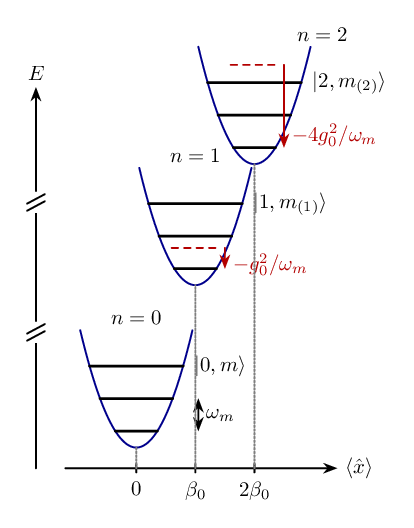}
\caption{Dressed-state level structure of the transformed Hamiltonian, Equation~(\ref{eq:hamiltonian_transformation}). Each cavity photon number $n$ carries its own mechanical oscillator manifold, displaced in equilibrium position by an amount proportional to $n\beta_0$ (dotted guides) and lowered in energy by the Kerr shift $-n^2 g_0^2/\omega_m$ relative to the undisplaced energy $n\omega_c$ (dashed levels and red arrows). The levels within each manifold are the displaced Fock states $\ket{n, m_{(n)}}$ of Equation~(\ref{eq:eigenstate_product}), spaced by $\omega_m$. The vertical separations between manifolds, of order $\omega_c$, are not to scale (axis breaks). Because the displacement depends on $n$, mechanical quanta exchanged with the reservoir carry which-path information about the photon number, which is the microscopic origin of the cavity dephasing discussed in Section~\ref{sec4}.}
\label{fig.ladder}
\end{figure}

A dressed state arises when the mechanical state becomes dependent on the photon number, so that the cavity photons effectively ``dress'' the mechanical state. The resulting effective photon-photon interaction, inherited from the underlying photon-phonon coupling, is quantum in origin and is often termed the intrinsic optomechanical non-linearity. In the single-photon regime considered here this non-linearity is subtle for small $\beta_0$ and becomes appreciable as $\beta_0 \to 1$. By construction of the polaron transformation, the dressed-state expectation of the original Hamiltonian equals the undressed expectation of $\hat{H}'$, namely $\langle n,m_{(n)}| \hat{H} | n,m_{(n)}\rangle = \langle n,m| \hat{H}' | n,m\rangle$. The photon-phonon basis $\ket{n,m}$ is taken as a tensor product of Fock states throughout. Figure~\ref{fig.ladder} summarizes the resulting level structure.

\section{Dressed-State Master Equation}\label{sec4}

In studying the dynamics of an open quantum system, a typical approach is to use the Standard Master Equation (SME) under the Born-Markov approximation. For systems with weak single-photon coupling, the photon-phonon interaction can be neglected, allowing us to approximate the system's eigenmodes and use the SME. However, in the strong single-photon coupling regime, intermodal interactions become significant and cannot be ignored, necessitating an alternative approach known as the Dressed-State Master Equation (DSME) \cite{Hu, Aga, Tah}. The name records the basis in which the equation is derived. The dissipators are obtained by decomposing the system-bath coupling in the eigenbasis of the interacting photon-phonon Hamiltonian, the dressed states of Section~\ref{sec3}, rather than in the eigenbasis of the two bare modes. The DSME is therefore formulated within the photon-phonon eigenbasis introduced in Section \ref{sec3}, where we consider the strong coupling between the cavity mode and the mechanical mode. In this basis, the Hamiltonian incorporates the shifts in energy levels caused by the photon-phonon interaction, producing dressed states that reflect the altered eigenstructure of the system. Additionally, it is assumed that the thermal bath coupled to the mechanical resonator interacts with both mechanical and optical degrees of freedom, while the bath coupled to the optical resonator affects only the optical degree of freedom.

In the interaction picture, the mechanical operator $\hat{b}(t)$ takes the form
\begin{equation}\label{6a}
\hat{b}(t) = e^{i H' t} \hat{b}\  e^{-i H' t},
\end{equation}
where $H'$ is given by Equation~(\ref{eq:hamiltonian_transformation}). Evaluating Equation~(\ref{6a}) by inserting the completeness relation for the dressed-state basis $\ket{n,l_{(n)}}$ from Equation~(\ref{eq:eigenstate_product}) isolates a zero-frequency component of the mechanical operator (see Appendix~\ref{C0} for the intermediate steps). The result is
\begin{equation}\label{6b}
\hat{b}(t) = e^{-i \omega_m t} (\hat{b} - \beta_0 \hat{N}_c) + \beta_0 \hat{N}_c.
\end{equation}

The term $\beta_0\hat{N}_c$ is the static displacement of the mechanical mode produced by the radiation pressure of $n$ cavity photons. It carries no time dependence because it shifts the energy levels without exchanging phonons with the reservoir, and therefore couples to the zero-frequency component of the bath spectral density. The term $e^{-i\omega_m t}(\hat{b} - \beta_0\hat{N}_c)$ describes quantum fluctuations around this static displacement and exchanges energy at frequency $\omega_m$. The same decomposition applied to the SME framework in the non-dressed photon--phonon basis $\ket{n,l}$ yields
\begin{equation}\label{6c}
\begin{split}
\hat{b}(t) &= e^{-i \omega_m t} \hat{b}\\
           &= e^{-i \omega_m t} (\hat{b}-\beta_0\hat{N}_c) + e^{-i \omega_m t}(\beta_0\hat{N}_c).
\end{split}
\end{equation}

It is evident from Equations~(\ref{6b}) and (\ref{6c}) that in the interaction picture for DSME, $\hat{b}(t)$ resolves into two terms rotating with different frequencies $\omega_m$ and zero. In Equation~(\ref{6b}), the first term changes the energy of the optomechanical system by exchanging one phonon with the mechanical thermal bath, while the second term shifts the system energy levels without exchanging any energy. The photon annihilation operator in the interaction picture, $\hat{a}(t)$, is similar for both DSME and SME at optical frequencies \cite{Hu},
\begin{equation}
\hat{a}(t) \approx e^{-i \omega_c t} \hat{a}.
\end{equation}

For a system in dressed basis (where $\beta_0 \approx 1$), the SME approach introduces additional term in the interaction picture,
\begin{equation}\label{7}
\begin{split}
\frac{d\Tilde{\rho} (t)}{dt} &\approx \gamma_m (n_{\text{th}} + 1) \mathcal{D}[\hat{b} - \beta_0 \hat{N}_c] \Tilde{\rho}(t) + \gamma_m n_{\text{th}} \mathcal{D}[\hat{b}^\dagger - \beta_0 \hat{N}_c] \Tilde{\rho}(t) \\
&\quad + \kappa \mathcal{D}[\hat{a}] \Tilde{\rho}(t) + \gamma_m (2 n_{\text{th}} + 1) \beta_0^2 \mathcal{D}[\hat{N}_c] \Tilde{\rho}(t),
\end{split}
\end{equation}
where $\Tilde{\rho}(t)$ is the density matrix in the interaction picture, $n_{\text{th}}$ is the thermal occupancy number at temperature $T$, $\gamma_m$ and $\kappa$ are the mechanical and optical decay rates, respectively, and $\hat{N}_c$ is the cavity photon number operator. The term $\mathcal{D}[\hat{o}]\Tilde{\rho}(t)$ represents the dissipator, defined here as
\begin{equation}\label{6}
\mathcal{D}[\hat{o}]\Tilde{\rho}(t) = \hat{o} \Tilde{\rho}(t) \hat{o}^\dagger - \frac{1}{2} \left\{ \hat{o}^\dagger \hat{o}, \Tilde{\rho}(t) \right\},
\end{equation}
where $\hat{o}$ is a system eigen-operator.

The last term in Equation~(\ref{7}) corresponds to dephasing in the density matrix $\Tilde{\rho}(t)$, as it does not affect photon or phonon populations directly but still contributes to the decoherence of $\Tilde{\rho}(t)$. For $\beta_0 \ll 1$, terms involving $\beta_0$ can be neglected. This reduces Equation~(\ref{7}) to SME in the unperturbed eigenbasis. Assuming Ohmic reservoirs the DSME in the interaction picture can be written as
\begin{equation}\label{5}
\begin{split}
\frac{d\Tilde{\rho} (t)}{dt} &= \gamma_m (n_{\text{th}} + 1) \mathcal{D}[\hat{B}_m] \Tilde{\rho}(t) + \gamma_m n_{\text{th}} \mathcal{D}[\hat{B}_m^\dagger] \Tilde{\rho}(t) + \kappa \mathcal{D}[\hat{a}] \Tilde{\rho}(t)\\
&+4\gamma_m\left(\frac{k_B T}{\omega_m}\right)\beta_0^2\mathcal{D}[\hat{N_c}]\Tilde{\rho}(t),
\end{split}
\end{equation}
where $\hat{B}_m = \hat{b}-\beta_0\hat{N}_c$. Note here that an additional approximation is applied to cavity photons in the DSME approach. The cavity frequency should be much larger than the mechanical frequency ($\omega_c \gg \omega_m$), an assumption generally valid for optical photons inside the cavity. In the case of microwave photons, however, this approximation may not hold, requiring the use of the Global Master Equation (GME) \cite{Tah}. Comparing Equations~(\ref{7}) and (\ref{5}) we see that the SME in the dressed basis carries an extra dephasing term that the DSME does not, but with a coefficient $(2 n_{\text{th}}+1)$ rather than $4 k_B T/\omega_m$. At high reservoir temperature, $n_{\text{th}} \to k_B T/\omega_m$, the DSME coefficient is the larger of the two. This is the physically meaningful dephasing rate that the SME under-counts at high temperature, as first noted in \cite{Hu}. At low reservoir temperature, the dephasing term in Equation~(\ref{5}) is negligible, and the DSME predicts a higher coherence than the SME. The remainder of the paper examines how this comparison changes for a squeezed thermal reservoir and how the squeezing parameters provide control over the dephasing rate.

\section{Dressed-State Master Equation for a Squeezed Thermal Reservoir}\label{sec5}

In open quantum systems, the non-unitary evolution of the state density matrix arises from interactions with an external reservoir. A well-established method to study system-reservoir interactions is through the Born-Markov approximation, where the coupling between the system and reservoir is assumed to be weak, and the reservoir correlation functions exhibit ``memory-less'' behavior. In other words, the characteristic time scale for interactions among reservoir components is much shorter than that of the system. Experimental studies typically involve vacuum or thermal reservoirs. However, non-thermal reservoirs, such as squeezed thermal reservoirs, are more challenging to construct experimentally but have shown promising theoretical applications. Squeezed thermal reservoirs, for instance, have been demonstrated to mitigate quantum coherence decay rates, thus aiding in the preservation of non-classical effects \cite{Ser, Kim}. Recent studies have highlighted the utility of squeezed thermal reservoirs as resources for enhancing the performance of nanomechanical quantum engines, allowing them to exceed the Carnot efficiency limit and improve both power output and efficiency in quantum heat engines \cite{Jan, Aba}. While there is extensive understanding of how squeezed reservoirs influence decoherence in individual quantum states (for example coherent or Fock states), there remains a significant gap in the literature regarding the effects of squeezed reservoirs on the decoherence properties of strongly coupled systems. Understanding this role becomes especially relevant for optomechanical systems operating in the strong photon-phonon coupling regime.

\subsection{Properties and Realization of Squeezed Thermal Reservoirs}\label{sec5a}

The squeezed reservoir entering the master equations below acts on the mechanical mode, so its realization requires engineering the phase-space statistics of the mechanical environment rather than of the optical field alone. The distinction matters because squeezed optical reservoirs are by now an established resource in cavity optomechanics. Injected squeezed light has been used to cool a mechanical oscillator below the quantum backaction limit \cite{Clark17}, to enhance measurement-based feedback cooling \cite{Scha16}, to suppress Stokes scattering in sideband cooling \cite{Asjad16}, and to dissipatively stabilize targeted Gaussian mechanical states in multimode settings \cite{Yazdi26}. An optical squeezed reservoir does not, however, translate by itself into a squeezed environment for the mechanics, and a dedicated engineering route is required for the mechanical bath assumed in this work.

A route that is both tunable and experimentally demonstrated is dissipative reservoir engineering with two-tone driving \cite{Kron13}. An auxiliary cavity mode of linewidth $\kappa_{\text{aux}}$ is driven at its two mechanical sidebands with effective linearized coupling amplitudes $G_-$ (red-detuned tone) and $G_+$ (blue-detuned tone). After adiabatic elimination of the driven mode, the mechanical mode relaxes as if coupled to a squeezed reservoir that damps the Bogoliubov operator $\hat{b}\cosh r_e + \hat{b}^\dagger \sinh r_e\, e^{i\theta}$, with squeezing strength set by the amplitude ratio through $\tanh r_e = G_+/G_-$ and phase $\theta$ set by the relative phase of the two drive tones \cite{Kron13}. Both reservoir parameters are therefore controlled by drive amplitudes and phases, which is precisely the tunability exploited in Section~\ref{sec6}. This mechanism has been realized independently in three microwave optomechanical experiments, each preparing mechanical states squeezed below the zero-point level \cite{Woll15, Pirk15, Leco15}. The engineered reservoir has a bandwidth of order $\kappa_{\text{aux}} \gg \gamma_m$, so the Born-Markov treatment used throughout this work remains valid, and the residual intrinsic thermal bath of the resonator adds to the engineered one so that the composite environment seen by the mechanics is a squeezed thermal reservoir of the form assumed below. The auxiliary mode operates in the standard linearized regime while the undriven primary mode carries the strong single-photon physics, so the two requirements do not compete. In particular, the two tones drive the auxiliary mode only and enter the description of the primary system solely through the reservoir parameters $N_{\text{eff}}$ and $M_{\text{eff}}$, so the undriven limit taken in Section~\ref{sec3} is unaffected by the reservoir engineering.

For the parameter regime simulated in Section~\ref{sec6}, namely $\beta_0 = 0.8$, $n_{\text{th}} = 20$ and an optimal squeezing strength $r \approx 1.4$, the two-tone route requires a sideband amplitude ratio $G_+/G_- = \tanh(1.4) \approx 0.89$, corresponding to approximately $12$~dB of effective reservoir squeezing. Maintaining the optimal $r$ amounts to stabilizing the ratio of two drive amplitudes, and maintaining $\theta = \pi$ amounts to phase locking the two tones, both routine operations in the microwave experiments cited above. Alternative routes exist. A modulated signal applied to a nanobeam resonator realizes squeezed thermal noise directly \cite{Jan}, interference between spontaneous decay channels in atomic systems yields an effective squeezed vacuum reservoir \cite{Zol}, and time-independent linear coupling of the system to an auxiliary lossy mode in an ordinary thermal environment has recently been shown to realize an effective squeezed thermal reservoir across circuit-QED and cavity-QED platforms without active modulation \cite{Lee2026}.

Squeezed thermal reservoirs are categorized as non-stationary reservoirs \cite{Hei}, in contrast to stationary thermal reservoirs. This classification arises because the correlation functions of the electric field operators for the radiation modes in squeezed reservoirs are time-dependent, breaking the time-homogeneity characteristic of thermal reservoirs. Consequently, in the rotating wave approximation, the master equation governing the evolution of optical and mechanical operators includes additional terms that account for this time-dependence. These modifications have significant implications for the decoherence behavior of systems interacting with squeezed thermal reservoirs, particularly under strong coupling conditions. We examine the Standard Master Equation (SME) in quantum optics for a system interacting with a squeezed thermal reservoir \cite{Hei}. In the context of our study, this corresponds to the SME in the unperturbed eigenbasis, or equivalently, the system's eigenbasis in the weak photon-phonon coupling regime ($\beta_0\ll 1$). The SME under these conditions is given by
\begin{equation}\label{8}
\begin{split}
\frac{d\Tilde{\rho} (t)}{dt}&=\gamma_m (N_{eff} + 1)\mathcal{D}[\hat{b}]\Tilde{\rho}(t)+\gamma_m(N_{eff})\mathcal{D}[\hat{b}^\dagger]\Tilde{\rho}(t)\\
&+\kappa (N+1)\mathcal{D}[\hat{a}]\Tilde{\rho}(t)+\kappa(N)\mathcal{D}[\hat{a}^\dagger]\Tilde{\rho}(t)\\
&-\gamma_m M_{eff}^*\hat{b}\Tilde{\rho}(t)\hat{b}-\gamma_m M_{eff}\hat{b}^\dagger \Tilde{\rho}(t)\hat{b}^\dagger\\
&-\kappa M^*\hat{a}\Tilde{\rho}(t)\hat{a}-\kappa M\hat{a}^\dagger\Tilde{\rho}(t)\hat{a}^\dagger.
\end{split}
\end{equation}
Equation~(\ref{8}) and the two dressed-basis master equations derived below, Equations~(\ref{13}) and (\ref{10}), are written in the compact $N$, $M$ notation and are to be read as shorthand for the manifestly Lindblad form obtained through the standard Bogoliubov transformation
\begin{equation}\label{eq:bogoliubov}
\begin{split}
    & \hat{C} = \hat{\beta}\cosh{r} + \hat{\beta}^\dagger\sinh{r}\exp{i\theta},\\
    & \hat{D} = \hat{\alpha}\cosh{r} + \hat{\alpha}^\dagger\sinh{r}\exp{i\theta},
    \end{split}
\end{equation}
where $\hat{\beta}$ and $\hat{\alpha}$ denote the relevant mechanical and optical jump operators, $\hat{b}$ and $\hat{a}$ in Equation~(\ref{8}) and $\hat{B}_m$ and $\hat{a}$ in the dressed-basis equations below. In terms of these operators the reservoir terms take the form $\gamma_m(n_{\text{th}}+1)\mathcal{D}[\hat{C}] + \gamma_m n_{\text{th}}\mathcal{D}[\hat{C}^\dagger]$ for the mechanics and $\kappa\,\mathcal{D}[\hat{D}]$ for the optical mode, which is also the form used in the numerical simulations of Section~\ref{sec6}. The terms $N_{\text{eff}}$ and $M_{\text{eff}}$ characterize the squeezed thermal reservoir and encapsulate the effects of thermal and squeezing parameters on the system dynamics. $M$ and $M_{\text{eff}}$ arise since the electric field operator for squeezed thermal reservoir is not homogeneous in time \cite{Hei}. Specifically, $N_{\text{eff}}$ and $M_{\text{eff}}$ are defined as follows:
\begin{equation}
N_{\text{eff}} = n_{\text{th}} (\cosh^2{r} + \sinh^2{r}) + \sinh^2{r},
\end{equation}
\begin{equation}
M_{\text{eff}} = -\cosh{r} \sinh{r} e^{i \theta} (2 n_{\text{th}} + 1),
\end{equation}
where $n_{\text{th}}$ represents the thermal occupancy of the mechanical mode, while $r$ and $\theta$ denote the squeezing parameters. Here, $r$ quantifies the degree of squeezing, and $\theta$ represents the phase of the squeezed quadrature. For optical photons, the thermal occupancy at the cavity frequency is negligible ($n_{\text{th}}(\omega_c) \approx 0$); thus, $N_{\text{eff}}$ and $M_{\text{eff}}$ reduce to simpler forms,
\begin{equation}
N = N_{\text{eff}} \big|_{n_{\text{th}} = 0} = \sinh^2{r},
\end{equation}
\begin{equation}
M = M_{\text{eff}} \big|_{n_{\text{th}} = 0} = -\cosh{r} \sinh{r} e^{i \theta},
\end{equation}
for the optical operator $\hat{a}$. These simplifications capture the effective noise parameters for the optical mode when it is subjected to a squeezed vacuum environment.

The terms proportional to $N_{\text{eff}}$ and $M_{\text{eff}}$ introduce unique contributions to the system's dissipative dynamics due to the non-classical properties of the squeezed thermal reservoir. In particular, the presence of $M_{\text{eff}}$ reflects the phase-sensitive correlations in the noise spectrum of the reservoir, a hallmark of squeezed states. Such correlations can lead to reduced decoherence rates for certain system states, thereby helping preserve quantum coherence. This characteristic has implications for applications that rely on long-lived quantum states, such as quantum information processing and precision sensing.

\subsection{SME for $\beta_0\approx 1$ under the influence of a squeezed thermal reservoir}\label{sec5b}
In Section \ref{sec4} we showed that for strong single-photon optomechanical coupling ($\beta_0\approx 1$), the SME in the interaction picture for a thermal reservoir contains an additional cavity dephasing term (Equation~(\ref{7})). For a squeezed thermal reservoir the same structure persists but with two extra terms, both arising from the time-dependent correlation functions of the squeezed bath. The derivation is given in Appendix \ref{C2} and yields
\begin{equation}\label{13}
\begin{aligned}
    \frac{d\Tilde{\rho} (t)}{dt}&=\gamma_m (N_{eff} + 1)\mathcal{D}[\hat{B}_m]\Tilde{\rho}(t) + \gamma_m N_{eff}\mathcal{D}[\hat{B}_m^\dagger]\Tilde{\rho}(t)\\
&+\kappa(N+1)\mathcal{D}[a]\Tilde{\rho}(t)+\kappa(N)\mathcal{D}[\hat{a}^\dagger]\Tilde{\rho}(t)\\
&-\gamma_m M_{eff}^*(\hat{B}_m)\Tilde{\rho}(t)(\hat{B}_m)-\gamma_m M_{eff}(\hat{B}_m)^\dagger \Tilde{\rho}(t)(\hat{B}_m)^\dagger\\
&-\kappa M^*\hat{a}\Tilde{\rho}(t)\hat{a}-\kappa M\hat{a}^\dagger\Tilde{\rho}(t)\hat{a}^\dagger\\
&+2\gamma_m\cosh{r}\sinh{r}\cos{\theta}(2n_{\text{th}}+1)\beta_0^2\mathcal{D}[\hat{N}_c]\Tilde{\rho}(t)\\
&+\gamma_m(2 N_{eff}+1)\beta_0^2\mathcal{D}[\hat{N_c}]\Tilde{\rho}(t).
\end{aligned}
\end{equation}
Comparing with Equation~(\ref{7}), in the last term $n_{\text{th}}$ is replaced by $N_{eff}$, and the second-last term in Equation~(\ref{13}) is proportional to the cosine of the reservoir phase. Both extra terms vanish in the weak photon-phonon coupling limit ($\beta_0\ll 1$).

\subsection{DSME for $\beta_0\approx 1$ under the influence of a squeezed thermal reservoir}\label{sec5c}
We now construct the corresponding DSME, which removes the spurious dephasing of the SME by resolving the distinct frequency components of $\hat{b}(t)$. Following the procedure of Hu \etal \cite{Hu}, we begin by writing down the correlation function for a squeezed thermal reservoir. Because the electric-field operator is no longer time-homogeneous, the correlation function acquires both time-independent and time-dependent contributions \cite{Hei}. Rearranging,
\begin{equation}\label{9}
\begin{aligned}
\mathcal{R}(s,t) &= \sum_k |G^\lambda_k|^2(N_k-M_k\exp(-2i\omega_k t))e^{i\omega_k s}\\
&+ |G^\lambda_k|^2((N_k+1)-M_k^*\exp(2i\omega_k t))e^{-i\omega_k s},
\end{aligned}
\end{equation}
where
\begin{equation}
\langle O_\lambda^\dagger(\vec{k}) O_{\lambda'}(\vec{k'}) \rangle = \delta_{\lambda \lambda'} \delta_{\vec{k} \vec{k'}} N_k,
\end{equation}
\begin{equation}
\langle O_\lambda(\vec{k}) O_{\lambda'}^\dagger(\vec{k'}) \rangle = \delta_{\lambda \lambda'} \delta_{\vec{k} \vec{k'}} (N_k + 1),
\end{equation}
\begin{equation}
\langle O_\lambda(\vec{k}) O_{\lambda'}(\vec{k'}) \rangle = \delta_{\lambda \lambda'} \delta_{\vec{k} \vec{k'}} M_k,
\end{equation}
\begin{equation}
\langle O_\lambda^\dagger(\vec{k}) O_{\lambda'}^\dagger(\vec{k'}) \rangle = \delta_{\lambda \lambda'} \delta_{\vec{k} \vec{k'}} M_k^*,
\end{equation}
and $G^\lambda_k$ is the system-reservoir coupling for the $k^{th}$ reservoir mode $\hat{O}$ with polarization $\lambda$. The correlation function in Equation~(\ref{9}) generates terms proportional to $\hat{a}$, $\hat{a}^\dagger$, $(\hat{b}-\beta_0\hat{N_c})$ and $(\hat{b}-\beta_0\hat{N_c})^\dagger$ in the dressed basis. Since $n_{\text{th}}(\omega_c)\approx 0$ for the cavity mode, $N_{k}$ and $M_{k}$ reduce to $N$ and $M$; for the mechanical mode they are $N_{eff}$ and $M_{eff}$.

Three elements of the derivation deserve emphasis, with the intermediate steps given in Appendix~\ref{C1}. First, the dressed basis enters through the decomposition of Equation~(\ref{6b}), which exists only in that basis. The rotating component $\hat{B}_m$ samples the reservoir spectral density at $\omega_m$ and produces the jump terms, while the static component $\beta_0\hat{N}_c$ samples the spectral density at zero frequency and produces pure cavity dephasing. Second, for a squeezed reservoir the zero-frequency noise power is set by the variance of the reservoir quadrature selected by the phase $\theta$, which evaluates to $(k_B T/\omega_m)(\cosh^2{r}+\sinh^2{r}+2\cosh{r}\sinh{r}\cos{\theta})$ as shown in Equation~(\ref{c4}). This is the origin of the phase controllability of the dephasing term, and it is a strong-coupling effect because the static component carries the factor $\beta_0^2$. Third, the phase-sensitive $M_{eff}$ terms survive the rotating wave approximation because the explicit time dependence $e^{\pm 2i\omega_m t}$ of the reservoir correlation function in Equation~(\ref{9}) cancels the corresponding rotation of the system operators. The derivation in Appendix \ref{C1} then gives
\begin{equation}\label{10}
\begin{aligned}
\frac{d\Tilde{\rho} (t)}{dt}&=\gamma_m (N_{eff} + 1)\mathcal{D}[\hat{B_m}]\Tilde{\rho}(t) + \gamma_m N_{eff}\mathcal{D}[\hat{B_m}^\dagger]\Tilde{\rho}(t)\\
&+\kappa(N+1)\mathcal{D}[a]\Tilde{\rho}(t)+\kappa(N)\mathcal{D}[\hat{a}^\dagger]\Tilde{\rho}(t)\\
&-\gamma_m M_{eff}^*\hat{B_m}\Tilde{\rho}(t)\hat{B_m}-\gamma_m M_{eff}\hat{B_m}^\dagger \Tilde{\rho}(t)\hat{B_m}^\dagger\\
&-\kappa M^*\hat{a}\Tilde{\rho}(t)\hat{a}-\kappa M\hat{a}^\dagger\Tilde{\rho}(t)\hat{a}^\dagger\\
&+4\gamma_m\left(\frac{k_B T}{\omega_m}\right)(\cosh^2{r}+\sinh^2{r}+2\cosh{r}\sinh{r}\cos{\theta})\beta_0^2\mathcal{D}[\hat{N_c}]\Tilde{\rho}(t),
\end{aligned}
\end{equation}
where $\hat{B}_m=\hat{b}-\beta_0\hat{N_c}$. As expected, Equation~(\ref{10}) is structurally Equation~(\ref{8}) with $\hat{b}$ replaced by $\hat{B}_m$. The final term is the analogue of the dephasing term in Equation~(\ref{5}), now with the temperature- and $\beta_0$-dependent coefficient further modulated by the squeezing parameters $r, \theta$. For $r=0$ Equation~(\ref{10}) reduces to Equation~(\ref{5}). For $r \neq 0$ the squeezing parameters provide a tunable control over the photon-dephasing contribution to the decoherence of $\Tilde{\rho}(t)$. In Section \ref{sec6} we exploit this control to demonstrate that decoherence rates of coherent superposition states can be reduced approximately sevenfold for appropriate values of $r$ and $\theta$.

\section{Results and discussion}\label{sec6}
In this section, we present a comparative analysis of the decoherence of the system's density operator for thermal and squeezed thermal reservoirs within the DSME framework. We examine how the squeezing parameters $r$ and $\theta$ control the decoherence rate of cavity photon Fock superpositions in the strong single-photon coupling regime. The simulations use the \texttt{mesolve} routine from the Quantum Toolbox in Python (QuTiP) \cite{Qut1} with the master equations of Equations~(\ref{5}), (\ref{10}) and (\ref{13}) implemented in the Lindblad form stated after Equation~(\ref{8}). Time is reported throughout in units of the mechanical period through the dimensionless combination $\omega_m t$.

The initial state of the optomechanical system is
\begin{equation}\label{14}
\ket{\psi_{c,m}(0)} = \left( \cos\left( \frac{\zeta}{2} \right) \ket{p} + e^{i\phi} \sin\left( \frac{\zeta}{2} \right) \ket{q} \right)_c \otimes \ket{u}_m,
\end{equation}
where the subscripts $c$ and $m$ denote the cavity photon and mechanical Fock states, respectively. As shown in Section~\ref{sec3}, these states are eigenstates of the transformed Hamiltonian $\hat{H}'$. Coherent-state superpositions would be experimentally more accessible and do not change the qualitative results, but we use Fock states for technical clarity. In the interaction picture the coherence of the cavity photon state is captured by the off-diagonal element of the reduced cavity density operator,
\begin{equation}\label{15}
P_{pq}(t) = \left| \bra{p} \tilde{\rho}_c(t) \ket{q} \right| = \left| \bra{p} \operatorname{Tr}_m\left[ \tilde{\rho}_{c,m}(t) \right] \ket{q} \right|,
\end{equation}
where $\tilde{\rho}_{c,m}(0) = \ket{\psi_{c,m}(0)}\bra{\psi_{c,m}(0)}$ is the initial density operator and $\operatorname{Tr}_m[\cdots]$ denotes the partial trace over the mechanical Fock states. The general parametrization of Equation~(\ref{14}) records that the analysis places no restriction on the weights or the relative phase of the initial superposition. For the remainder of this section we fix $\zeta/2 = \pi/4$, $\phi=0$, $\ket{p}=\ket{0}$, and $\ket{q}=\ket{3}$. The equal-weight choice $\zeta = \pi/2$ maximizes the initial coherence, $P_{pq}(0) = 1/2$, and this state mirrors the one used in Hu \etal \cite{Hu}, allowing direct comparison with their results.

\begin{figure}[ht]
\centering
\includegraphics[width=0.6\linewidth]{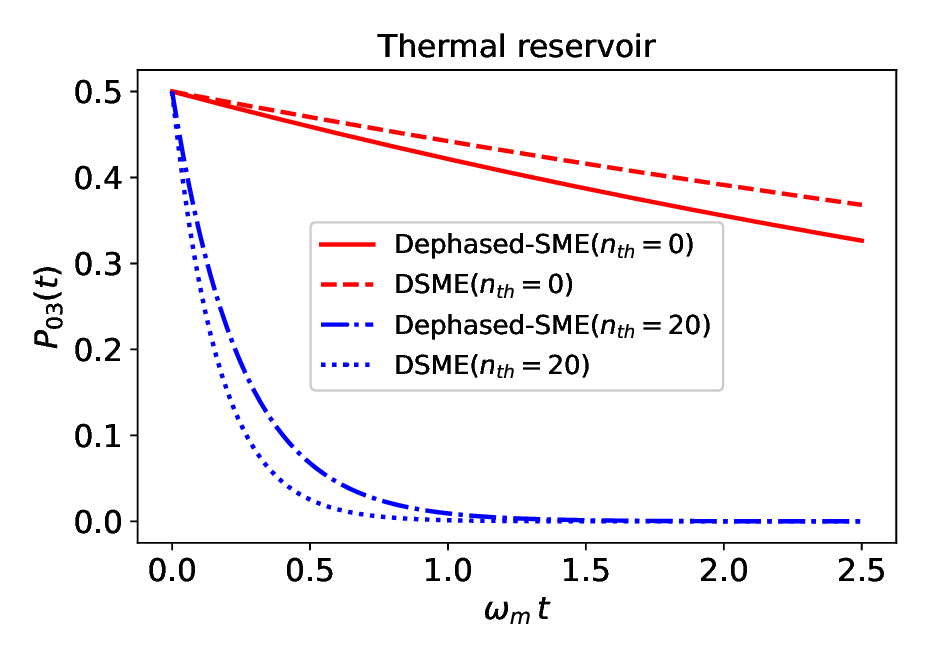}
\caption{Evolution of the off-diagonal element $|\bra{0}\operatorname{Tr}_m[\Tilde{\rho}_{c,m}(t)]\ket{3}|$ in a vacuum and thermal reservoir at $n_{\text{th}}=0$ and $n_{\text{th}}=20$. The optomechanical system parameters are $\beta_0=0.8$, $\kappa=0.05\,\omega_m$, and $\gamma_m=\kappa/3$.}
\label{fig.1}
\end{figure}

\begin{figure}[ht]
\centering
\begin{tabular}{cc}
\includegraphics[width=0.45\linewidth]{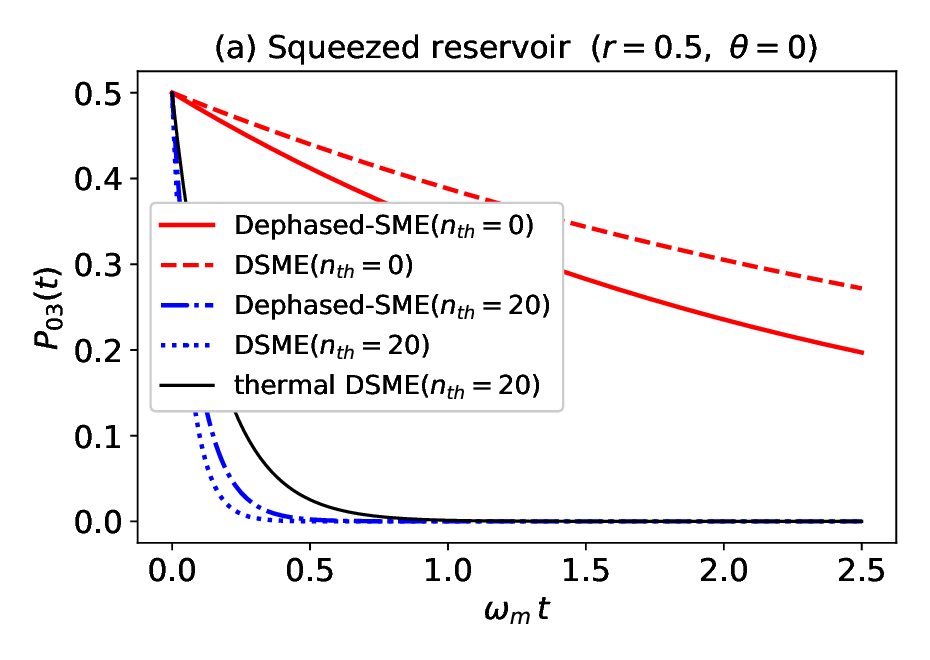} &
\includegraphics[width=0.45\linewidth]{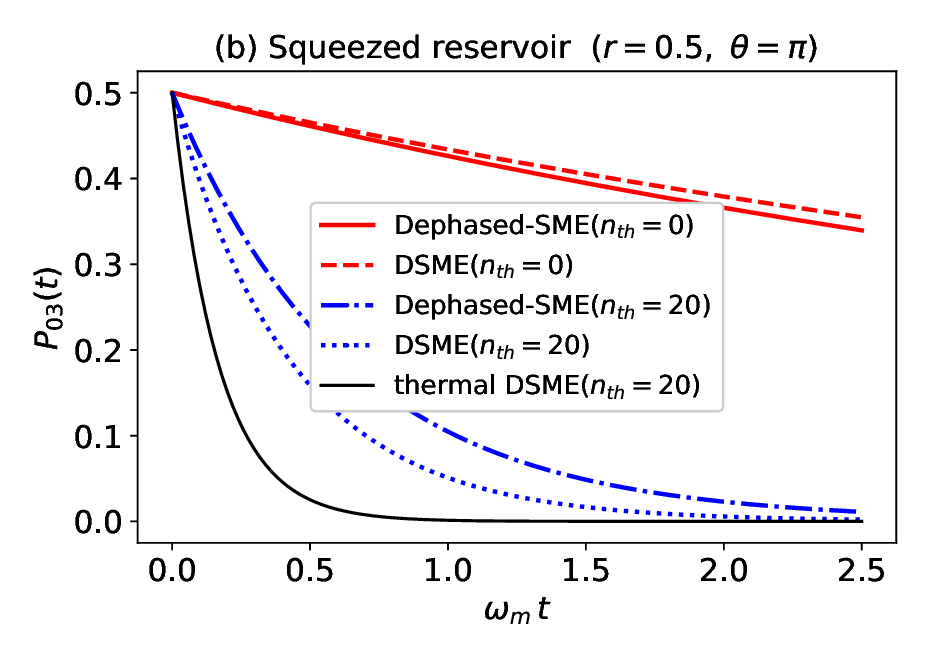}
\end{tabular}
\caption{Evolution of the off-diagonal element $|\bra{0}\operatorname{Tr}_m[\Tilde{\rho}_{c,m}(t)]\ket{3}|$ in a squeezed vacuum/thermal reservoir with $r=0.5$, at $n_{\text{th}}=0$ and $n_{\text{th}}=20$, for (a) reservoir phase $\theta=0$ and (b) $\theta=\pi$. The solid black curve reproduces the thermal DSME result of Figure~\ref{fig.1} at $n_{\text{th}}=20$ \cite{Hu} for direct comparison. The optomechanical system parameters are $\beta_0=0.8$, $\kappa=0.05\,\omega_m$, and $\gamma_m=\kappa/3$.}
\label{fig.2}
\end{figure}

Figure~\ref{fig.1} shows the evolution of $P_{03}$ as a function of dimensionless time $\omega_m t$ for dephased-SME and DSME in vacuum and thermal reservoirs, corresponding to the master equations in Equations~(\ref{7}) and (\ref{5}). The behaviour is qualitatively similar to Hu \etal \cite{Hu}: the effect of cavity dephasing dominates at high temperatures in the DSME more strongly than in the SME in the interaction picture, because the DSME captures the full physical dephasing rate that the SME under-counts. A similar trend appears in Figure~\ref{fig.2}, where the reservoir is squeezed. Figure~\ref{fig.2}(a) ($\theta=0$) shows that the gap between SME and DSME at high reservoir temperature ($n_{\text{th}}=20$) is smaller than in Figure~\ref{fig.1}, while Figure~\ref{fig.2}(b) ($\theta=\pi$) shows the opposite effect.

At low reservoir temperatures ($n_{\text{th}}=0$), $P_{03}$ decays more slowly under DSME than under SME. The DSME at low temperature has no temperature-dependent dephasing (Equation~(\ref{10}) $\big|_{T \to 0}$), whereas the last term of the SME in Equation~(\ref{13}) does not vanish and contributes additional dephasing even at $n_{\text{th}}=0$.

Figures~\ref{fig.1} and \ref{fig.2} differ only in the reservoir. Figure~\ref{fig.1} is the thermal case $r=0$, while the two panels of Figure~\ref{fig.2} show the same system coupled to a squeezed reservoir with $r=0.5$ at the two phases $\theta=0$ and $\theta=\pi$. The comparison makes the meaning of reservoir control precise. Once the temperature is fixed, a thermal reservoir has no remaining free parameter, so its decoherence rate is fixed, as the identical black reference curve in both panels of Figure~\ref{fig.2} illustrates. A squeezed thermal reservoir at the same temperature retains two free parameters, and varying them moves the decay curves to either side of the thermal reference. The phase-controlled coherence enhancement in weakly coupled systems is established in the literature; the new finding here is that in the strong single-photon coupling regime ($\beta_0\approx 1$) the same phase control becomes a quantitatively dominant effect, captured explicitly by the $\cos\theta$-dependent term in Equations~(\ref{10}) and (\ref{13}).

\begin{figure}[ht]
\centering
\begin{tabular}{cc}
\includegraphics[width=0.45\linewidth]{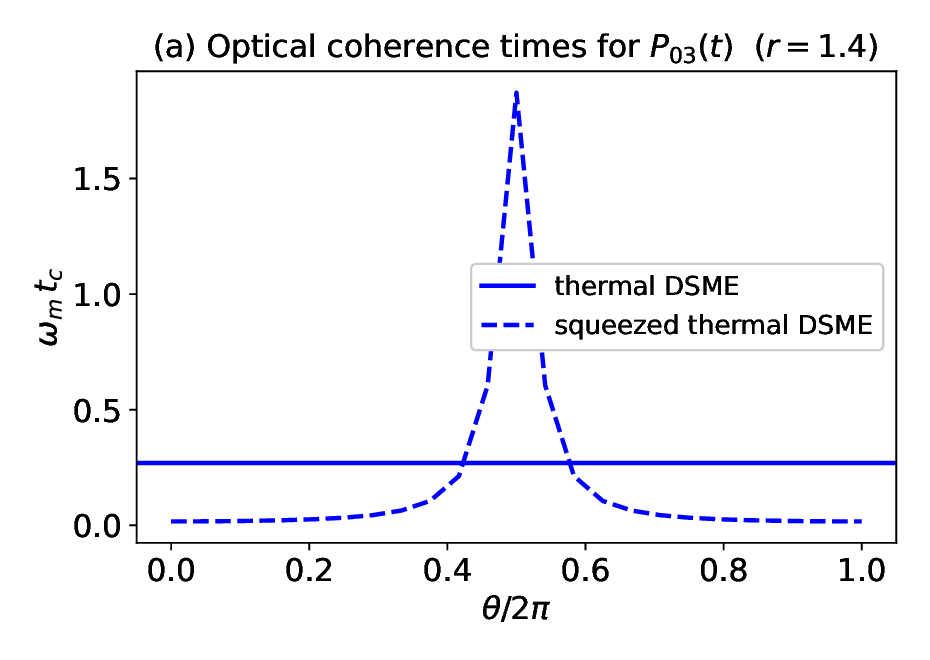} &
\includegraphics[width=0.45\linewidth]{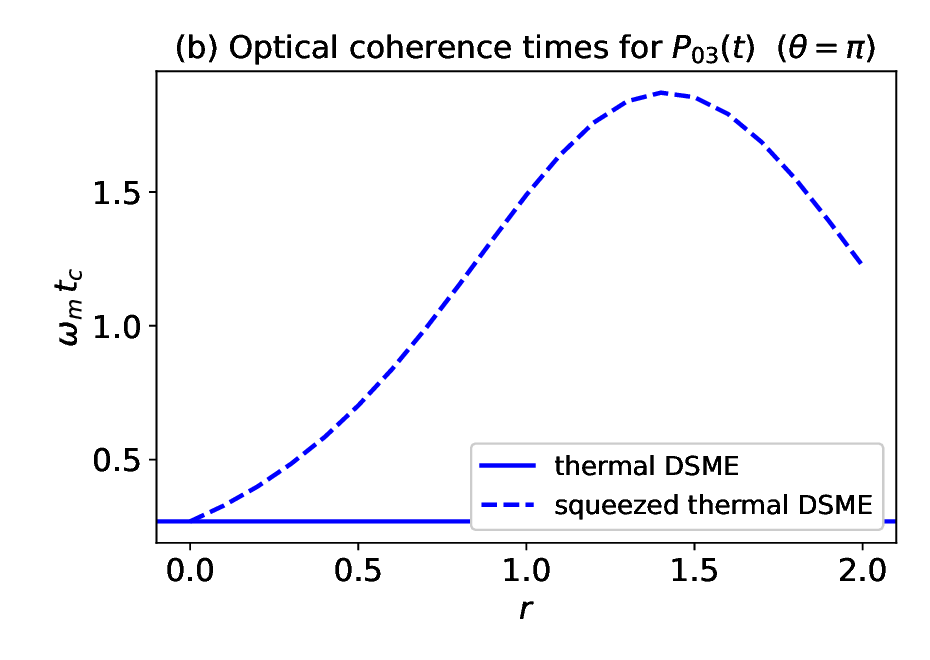}
\end{tabular}
\caption{Dimensionless coherence time $\omega_m t_c$, defined as the time at which $P_{03}$ falls below $0.1$, that is $20\%$ of its initial value $0.5$, under the influence of thermal and squeezed thermal reservoirs ($n_{\text{th}}=20$). Panel (a) varies the reservoir phase at fixed $r=1.4$ and panel (b) varies the squeezing strength at fixed $\theta=\pi$. The optomechanical system parameters are $\beta_0=0.8$, $\kappa=0.05\,\omega_m$, and $\gamma_m=\kappa/3$.}
\label{fig.3}
\end{figure}

For the same value of $r$, the decoherence rate changes substantially with $\theta$ in Figure~\ref{fig.2}, reflecting the $\cos\theta$ dependence of the dephasing term in Equation~(\ref{10}). In the weak-coupling limit ($\beta_0\ll 1$) the $\beta_0$-dependent terms in Equations~(\ref{10}) and (\ref{13}) vanish and no such explicit phase dependence appears. Figure~\ref{fig.3} characterises the dependence of the DSME coherence time on $r$ and $\theta$ in the strong-coupling regime. We denote by $t_c$ the time at which $P_{03}$ falls below $0.1$, that is $20\%$ of its initial value $0.5$, and use the dimensionless combination $\omega_m t_c$ as a proxy for the coherence time. The subscript distinguishes this fixed threshold-crossing time from the running time variable $t$ of the preceding figures.

The thermal-reservoir coherence time (solid blue line) is independent of $r$ and $\theta$, as expected. For the squeezed thermal reservoir, Figure~\ref{fig.3}(a) shows that the coherence time is maximal at the reservoir phase $\theta=\pi$, and Figure~\ref{fig.3}(b) locates the optimal squeezing strength near $r \approx 1.4$ at $\beta_0=0.8$, where the coherence time $\omega_m t_c \approx 1.9$ is approximately seven times the thermal-reservoir value $\omega_m t_c \approx 0.27$ at the same temperature. Beyond this optimum the coherence time decreases again, even with the reservoir phase held at $\theta=\pi$.

The non-monotonicity reflects a competition between terms in the DSME with opposite dependence on the squeezing strength. At $\theta=\pi$ the dephasing coefficient in Equation~(\ref{10}) reduces to $\cosh(2r)-\sinh(2r)=e^{-2r}$, which is suppressed exponentially with $r$. This term couples to the zero-frequency spectral density $J(0)$ and by itself would predict a coherence time that grows without bound as $r$ increases. Opposing it are the jump contributions---those proportional to $N_{\text{eff}}$ and $M_{\text{eff}}$ in Equation~(\ref{10})---whose prefactors grow as $\sim e^{2r}$ for $n_{\text{th}}\gg 1$ and couple to $J(\omega_m)$. The coherence time therefore peaks at the $r$ where the decreasing dephasing contribution and the growing jump contribution balance. Section~\ref{sec7} provides a quantitative scaling estimate that confirms this competition is a structural consequence of the DSME and not an artefact of the chosen numerical parameters.

\subsection{Phase-space picture}\label{sec6a}

\begin{figure}[ht]
\centering
\includegraphics[width=0.95\linewidth]{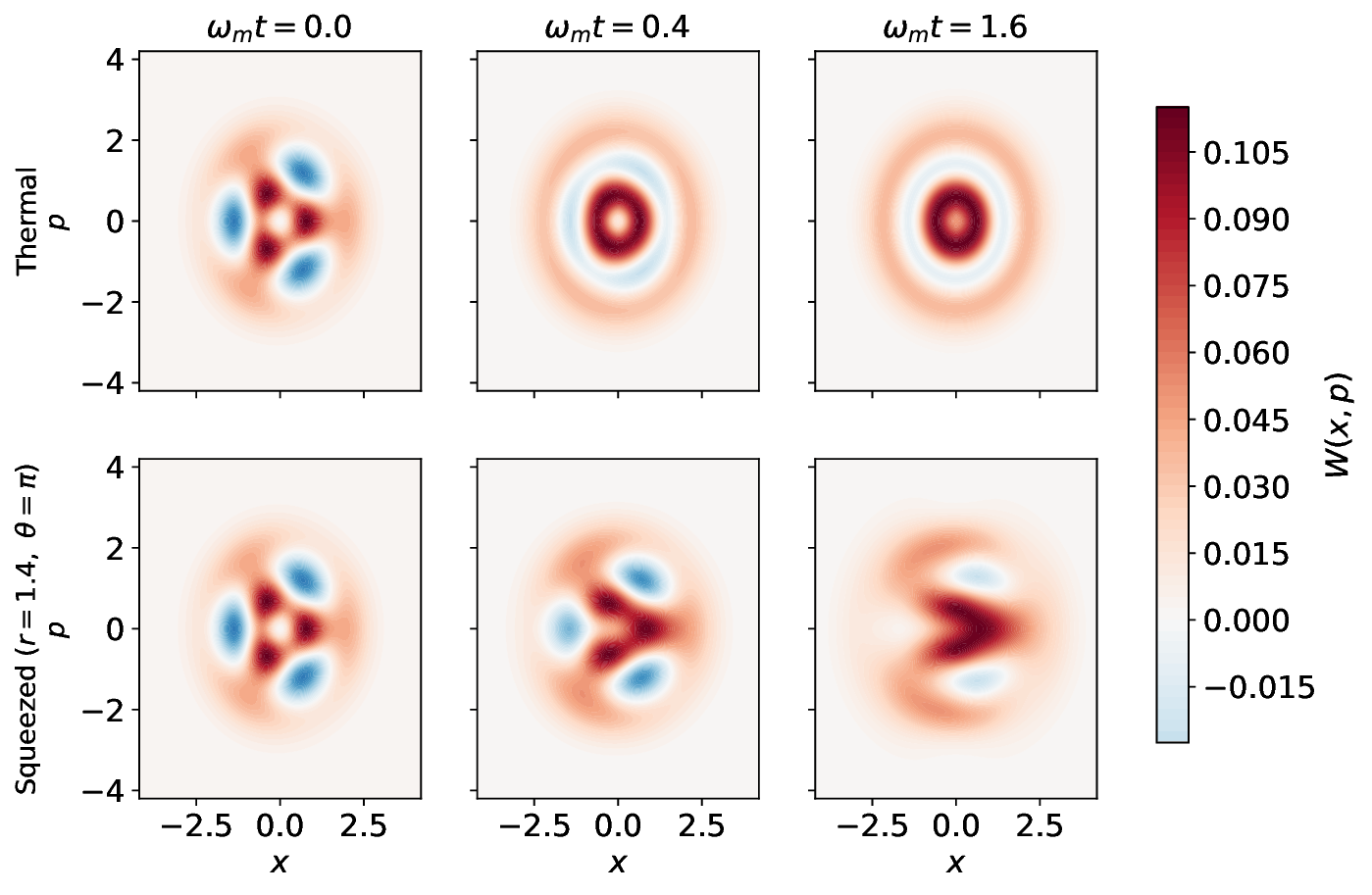}
\caption{Wigner function $W(x,p)$ of the reduced cavity state at three times, for the thermal reservoir (top row) and the squeezed thermal reservoir at the optimum $r=1.4$, $\theta=\pi$ (bottom row), both at $n_{\text{th}}=20$ and evolved under the DSME. Red denotes positive and blue negative values of $W$. The optomechanical system parameters are those of Figure~\ref{fig.1}.}
\label{fig.wigner}
\end{figure}

The phase-space consequences of this control are shown in Figure~\ref{fig.wigner}, which plots the Wigner function of the reduced cavity state at three times for the thermal reservoir and for the squeezed reservoir at the optimal point $r=1.4$, $\theta=\pi$ of Figure~\ref{fig.3}. At $t=0$ the state carries the threefold interference structure characteristic of the $\ket{0}$, $\ket{3}$ superposition. Under the thermal reservoir the interference lobes are erased by $\omega_m t \approx 0.4$, consistent with the coherence time read off Figure~\ref{fig.3}, and the state approaches a phase-insensitive mixture whose residual negative rings are those of the Fock component $\ket{3}$. Under the squeezed reservoir the interference structure survives essentially intact over the same interval and is still discernible at $\omega_m t = 1.6$, well beyond the thermal decoherence time.

\begin{figure}[ht]
\centering
\includegraphics[width=0.88\linewidth]{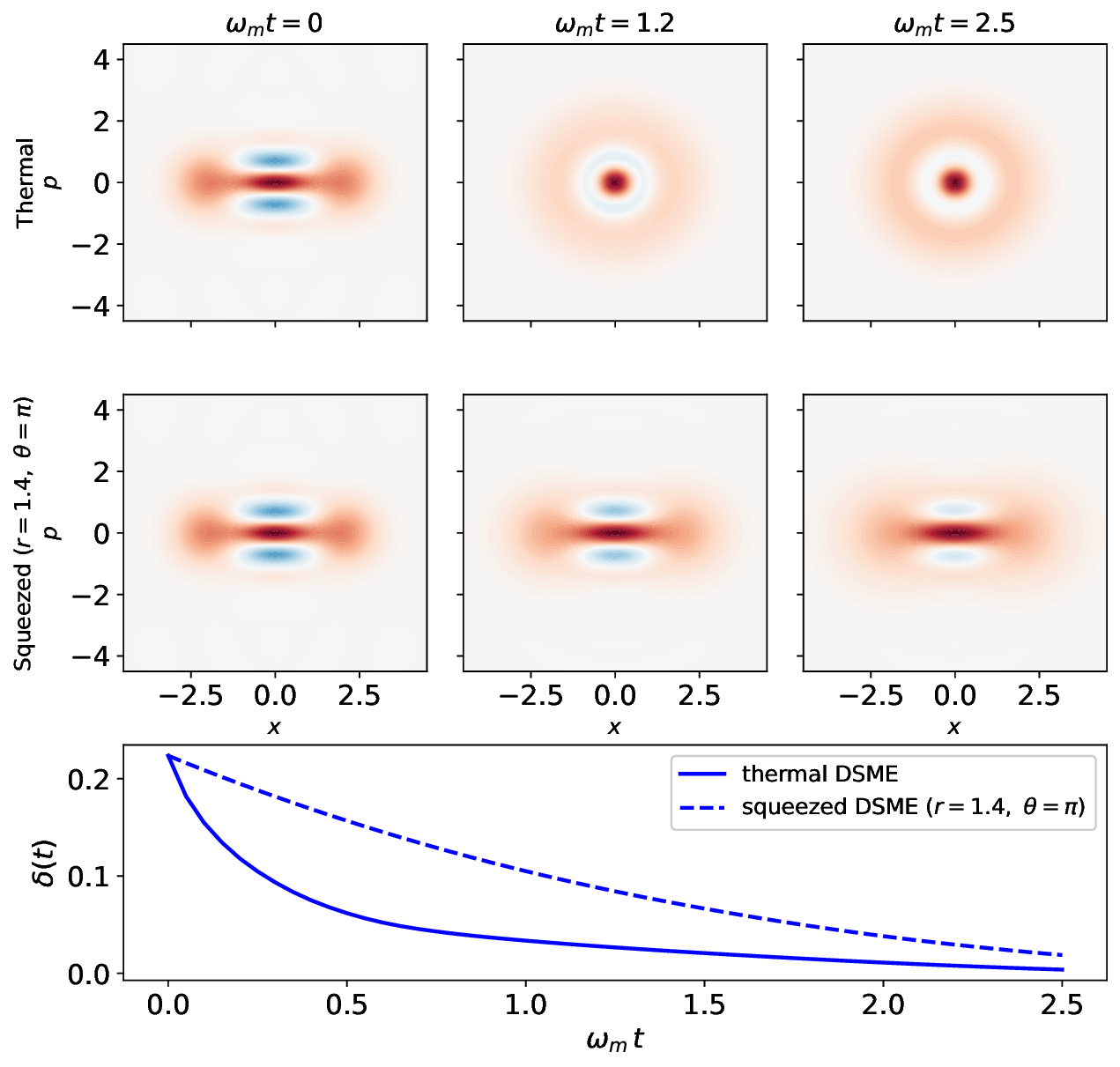}
\caption{Cat-state dynamics under the two reservoirs. Top and middle rows, Wigner function of an even coherent-state superposition with $\alpha=\sqrt{2}$ at three times under the thermal reservoir and under the optimally squeezed reservoir ($r=1.4$, $\theta=\pi$), both at $n_{\text{th}}=20$ and evolved under the DSME. Bottom row, the corresponding Wigner negativity volume $\delta(t)$. The thermal reservoir erases the interference fringes, while under the squeezed reservoir the fringe pattern remains resolved and its negativity decays several times more slowly. The optomechanical system parameters are those of Figure~\ref{fig.1}.}
\label{fig.neg}
\end{figure}

Because a dephased Fock mixture retains the ring negativity of its Fock components, the total negativity volume of the states in Figure~\ref{fig.wigner} is not by itself a faithful witness of the superposition coherence. A coherent-state superposition provides a cleaner witness, since the incoherent mixture of two coherent states has a positive Wigner function and all negativity is carried by the interference fringes. Figure~\ref{fig.neg} therefore shows the Wigner function of an even cat state $\propto \ket{\alpha} + \ket{-\alpha}$ with $\alpha=\sqrt{2}$ at three times, together with its negativity volume $\delta(t) = \frac{1}{2}\int \left[\,|W(x,p)| - W(x,p)\,\right] dx\, dp$, evolving under the same two reservoirs. The snapshots make the contrast visible. The thermal reservoir erases the fringes within a fraction of the plotted window while leaving the two coherent lobes, whereas under the squeezed reservoir the fringe pattern, reshaped along the squeezed quadrature, remains resolved well beyond the thermal decoherence time. The negativity volume quantifies this. Both reservoirs degrade $\delta(t)$, but the squeezed reservoir does so substantially more slowly, retaining $40\%$ of the initial negativity at $\omega_m t = 1.2$ where the thermal reservoir retains $12\%$, and roughly five times more at $\omega_m t = 2.5$. These values are converged with respect to the cavity and mechanical Hilbert-space truncations and the phase-space integration grid. The cat-state dynamics confirms that the coherence protection demonstrated above for Fock superpositions carries over to the fringe negativity of coherent-state superpositions, which is the resource carried by bosonic cat code states.

\subsection{Significance of the absolute timescales}\label{sec6b}

In the units used here the coherence times of Figure~\ref{fig.3} are short in absolute terms. Even at the optimal squeezing the superposition survives for about a third of a mechanical period, and under the thermal reservoir for only a few percent of one. It is worth stating precisely what this does and does not imply. The absolute scale is set by the benchmark parameters inherited from Hu \etal \cite{Hu}, namely a mechanical quality factor $Q_m = \omega_m/\gamma_m = 60$ and $n_{\text{th}} = 20$, chosen to make decoherence visible within a short simulation window rather than to represent a good device. Both the dephasing coefficient $A$ and the jump coefficient $B$ of Equation~(\ref{eq:total_rate}) below are proportional to $\gamma_m$, so at fixed $\kappa$ the coherence time scales as $1/\gamma_m$ while the optimal squeezing strength of Equation~(\ref{eq:optimal_r}) is independent of $\gamma_m$. At a modest $Q_m = 10^4$ the same dynamics gives $\omega_m t_c$ of several hundred, that is tens of mechanical periods, with the same sevenfold squeezing gain on top. The gain is multiplicative and transfers unchanged to realistic hardware. As $\gamma_m$ improves the coherence time eventually saturates at the optical-loss limit of order $1/\kappa$, and the value $\kappa = 0.05\,\omega_m$ used here is itself a conservative choice.

The relevant clock for the strong-coupling regime is in any case not the mechanical period but the nonlinear interaction time. The Kerr term of Equation~(\ref{eq:hamiltonian_transformation}) accumulates relative phase between $\ket{0}$ and $\ket{3}$ at the rate $(q^2-p^2)\,\beta_0^2\,\omega_m = 5.76\,\omega_m$, so one $\pi$ nonlinear phase shift takes $\omega_m t \approx 0.55$. Against this clock the thermal coherence time $\omega_m t_c \approx 0.27$ means that the coherence dies before a single nonlinear operation completes, whereas the squeezed optimum $\omega_m t_c \approx 1.9$ accommodates roughly three. At the benchmark parameters the squeezed reservoir is therefore the difference between an unusable and a usable intrinsic nonlinearity at the same temperature. Protocols that rely on the mechanical revival at $\omega_m t = 2\pi$ remain out of reach at this benchmark by a factor of about three, a gap that any improvement in $Q_m$ or reduction in $n_{\text{th}}$ closes.

\section{Scaling estimate of the optimal squeezing strength}\label{sec7}
The non-monotonic dependence of the coherence time on $r$ at $\theta=\pi$ (Figure~\ref{fig.3}) can be understood from the opposing $r$-scaling of the dephasing and jump contributions to the Liouvillian. The argument below isolates the leading $r$-dependence of each contribution and shows that a finite optimum is structurally guaranteed.

The dephasing term in Equation~(\ref{10}) acts on a cavity coherence $|p\rangle\langle q|$ through
\begin{equation}\label{eq:diss_Nc}
\mathcal{D}[\hat{N}_c](|p\rangle\langle q|) = -\tfrac{1}{2}(p-q)^2\,|p\rangle\langle q|,
\end{equation}
where $p$ and $q$ are the cavity photon Fock-state indices appearing in the initial superposition of Equation~(\ref{14}). Equation~(\ref{eq:diss_Nc}) follows from the definition of the dissipator, $\mathcal{D}[\hat{o}]\rho = \hat{o}\rho\hat{o}^\dagger - \tfrac{1}{2}\{\hat{o}^\dagger\hat{o},\rho\}$, together with $\hat{N}_c|n\rangle = n|n\rangle$:
$\hat{N}_c|p\rangle\langle q|\hat{N}_c = pq|p\rangle\langle q|$ and $\tfrac{1}{2}\{\hat{N}_c^2,\,|p\rangle\langle q|\} = \tfrac{1}{2}(p^2+q^2)|p\rangle\langle q|$, whose difference is $-\tfrac{1}{2}(p-q)^2|p\rangle\langle q|$. The dephasing contribution to the decay rate of $P_{pq}(t) = |\langle p|\operatorname{Tr}_m[\Tilde{\rho}(t)]|q\rangle|$ is therefore
\begin{equation}\label{eq:gamma_dephase}
\Gamma_{\text{deph}}(r,\theta) = 2(p-q)^2\gamma_m\Bigl(\frac{k_B T}{\omega_m}\Bigr)\beta_0^2\,\bigl[\cosh(2r) + \sinh(2r)\cos\theta\bigr].
\end{equation}
At $\theta = \pi$, the bracket reduces to $\cosh(2r) - \sinh(2r) = e^{-2r}$, giving
\begin{equation}\label{eq:gamma_dephase_pi}
\Gamma_{\text{deph}}(r,\pi) = 2(p-q)^2\gamma_m\Bigl(\frac{k_B T}{\omega_m}\Bigr)\beta_0^2\,e^{-2r}.
\end{equation}
This contribution is suppressed exponentially with $r$ and, taken alone, would predict a coherence time that increases without bound.

The jump terms in Equation~(\ref{10})---the dissipators $\mathcal{D}[\hat{B}_m]$ and $\mathcal{D}[\hat{B}_m^\dagger]$ and the $M_{\text{eff}}$ terms---all carry prefactors that grow with $r$. For $n_{\text{th}} \gg 1$, the dominant $r$-dependence enters through
\begin{equation}\label{eq:Neff_growth}
N_{\text{eff}} = n_{\text{th}}\cosh(2r) + \sinh^2 r \;\sim\; \frac{n_{\text{th}}}{2}e^{2r}\quad (r \gtrsim 1),
\end{equation}
and $|M_{\text{eff}}| = \cosh r\sinh r\,(2n_{\text{th}}+1) \sim n_{\text{th}} e^{2r}/2$. These operators couple to the spectral density at $\omega_m$ and induce transitions in the mechanical subspace that, through the partial trace, contribute to the decay of $P_{pq}(t)$. The exact prefactor of the jump contribution depends on mechanical overlap matrix elements and cannot be written in closed form without solving the full mechanical dynamics. Its scaling with $r$, however, is fixed: the jump-induced contribution grows as $\sim e^{2r}$.

The total effective rate governing the coherence decay may therefore be written
\begin{equation}\label{eq:total_rate}
\Gamma(r) \approx A\,e^{-2r} + B\,e^{2r} + \Gamma_0,
\end{equation}
where $A = 2(p-q)^2\gamma_m(k_B T/\omega_m)\beta_0^2$ is fixed by Equation~(\ref{eq:gamma_dephase_pi}), $B$ collects the prefactors of the jump contributions and scales as $\sim \gamma_m n_{\text{th}}$, and $\Gamma_0$ contains $r$-independent terms (the optical loss $\kappa$ and the $n_{\text{th}}$-independent part of the mechanical dissipation). The symbol $\Gamma_0$ is chosen to avoid a clash with the Bogoliubov operator $\hat{C}$ of Equation~(\ref{eq:bogoliubov}). The competition between the decreasing first term and the growing second term guarantees a minimum of $\Gamma(r)$ at
\begin{equation}\label{eq:optimal_r}
r^* = \frac{1}{4}\ln\frac{A}{B}.
\end{equation}
For the parameters of Figure~\ref{fig.3} ($p=0$, $q=3$, $\beta_0=0.8$, $n_{\text{th}}=20$, $k_B T/\omega_m \approx n_{\text{th}}$ in the high-temperature limit), $A \approx 230\,\gamma_m$. The numerical optimum $r^* \approx 1.4$ then implies $B \approx A/e^{4r^*} \approx 0.9\,\gamma_m$, which is of order $\gamma_m$ as expected for a mechanical jump rate. The estimate confirms that the observed non-monotonicity is a robust structural prediction of the DSME: any parameter regime with $A/B \gg 1$ will exhibit a coherence-time maximum at finite $r$.

\begin{figure}[ht]
\centering
\includegraphics[width=0.62\linewidth]{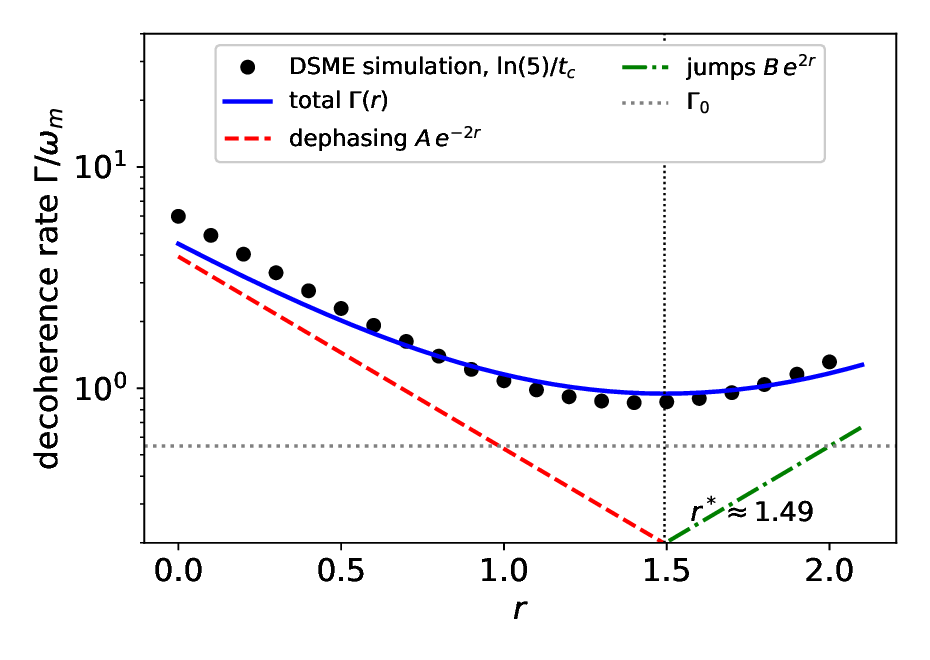}
\caption{Decomposition of the coherence decay rate at $\theta=\pi$. Points are the rates $\ln(5)/t_c$ obtained from the DSME simulations of Figure~\ref{fig.3}(b). The solid curve is Equation~(\ref{eq:total_rate}) with the dephasing coefficient $A$ fixed at the theoretical value of Equation~(\ref{eq:gamma_dephase_pi}) and with $B$ and $\Gamma_0$ fitted, and the broken curves show the three components separately. The competition between the falling dephasing term and the rising jump term produces the rate minimum, and hence the coherence-time maximum, near $r^* \approx 1.4$.}
\label{fig.rates}
\end{figure}

Figure~\ref{fig.rates} tests this estimate directly against the simulations. The decoherence rates $\Gamma = \ln(5)/t_c$ extracted from the coherence times of Figure~\ref{fig.3}(b) are compared with Equation~(\ref{eq:total_rate}), with $A$ fixed at its theoretical value and $B$ and $\Gamma_0$ fitted. The two-exponential model reproduces the simulated rates across the full range of $r$, and the fitted jump coefficient $B \approx 0.6\,\gamma_m$ places the predicted optimum at $r^* = \frac{1}{4}\ln(A/B) \approx 1.5$, in reasonable agreement with the observed maximum near $r \approx 1.4$.

\section{Conclusion}\label{sec8}
Three results follow from the analysis. First, the DSME for an optomechanical system in the strong single-photon coupling regime acquires, under a squeezed thermal reservoir, an additional dephasing term proportional to the cosine of the reservoir phase $\theta$. This term has no analogue in the weak-coupling limit and is the mechanism through which the squeezing parameters of the reservoir act as a control on the system's coherence. Second, setting the reservoir phase to $\theta=\pi$ at fixed squeezing strength reduces the dephasing rate substantially, extending the coherence time of cavity photon Fock superpositions approximately sevenfold relative to a thermal reservoir at the same temperature. Third, the dependence on the squeezing strength $r$ is non-monotonic: the coherence time peaks near $r\approx 1.4$ and decreases again for larger $r$. This peak arises from a competition between the dephasing term, whose coefficient at $\theta=\pi$ simplifies to $e^{-2r}$ and decreases with $r$, and the jump terms, whose prefactors grow as $N_{\text{eff}}\sim e^{2r}$ (Section~\ref{sec7}).

These findings position phase-sensitive reservoir engineering as a quantitative coherence-preservation tool in the strong-coupling regime that has no quantitatively comparable counterpart for thermal reservoirs. The phase-space analysis of Section~\ref{sec6a} shows the same control at the level of the Wigner function, including a severalfold slower decay of the fringe negativity of coherent-state superpositions. A natural extension is the entanglement dynamics of strongly coupled systems under phase-sensitive reservoirs.

Current bare optomechanical platforms have not yet reached single-photon couplings $\beta_0 \approx 1$ with simultaneously resolved sidebands ($\kappa \ll \omega_m$), which is the regime assumed here. The raw ratio $g_0/\omega_m \gtrsim 1$ has been demonstrated in sliced photonic-crystal nanobeams \cite{Leij15} and in nanowire-in-microcavity devices \cite{Fogl21}, but at the cost of optical linewidths far exceeding the coupling, and recent hybrid architectures are closing the remaining gap. Cavity exciton-polariton optomechanical systems have recently demonstrated single-particle resonant couplings approaching the ultrastrong regime, with the polariton Rabi gap protecting the coupling from inhomogeneous broadening \cite{Sesin2023}. As such platforms continue to push $\beta_0$ toward unity, the analysis presented here becomes directly applicable to coherence-preservation engineering, and the explicit phase dependence of the dephasing term in Equation~(\ref{10}) provides a concrete handle that designers of such systems can exploit. On the reservoir side, the two-tone dissipative engineering discussed in Section~\ref{sec5a} provides the tunable squeezed mechanical environment that the present analysis assumes, with the squeezing strength and phase set by drive amplitude ratios and relative phases.

\section*{Acknowledgements}
The authors acknowledge the use of Claude (Anthropic, version 4.7) partly for assistance with mathematical derivations and manuscript refinement. All AI-generated content was critically reviewed and verified by the authors. The authors received no financial support for this research. The authors declare no conflicts of interest.

\section*{Data availability statement}
The data that support the findings of this study are available from the corresponding author upon reasonable request.

\appendix
\section{Interaction-picture decomposition of $\hat{b}(t)$ in the dressed basis}\label{C0}
We derive Equation~(\ref{6b}) from the definition $\hat{b}(t) = e^{iHt}\,\hat{b}\,e^{-iHt}$, where $H$ is the original lab-frame Hamiltonian whose eigenstates are the dressed states $\ket{n,l_{(n)}}$ with eigenvalues $E_{n,l} = n\omega_c + l\omega_m - n^2g_0^2/\omega_m$. Insert two resolutions of the identity in this eigenbasis:
\begin{equation}\label{c0a}
\begin{aligned}
\hat{b}(t) &= \sum_{n,l}\sum_{n',l'} \ket{n,l_{(n)}}\bra{n,l_{(n)}} e^{iHt}\,\hat{b}\,e^{-iHt} \ket{n',l'_{(n')}}\bra{n',l'_{(n')}}\\
&= \sum_{n,l}\sum_{n',l'} e^{i(E_{n,l}-E_{n',l'})t}\; \ket{n,l_{(n)}}\bra{n,l_{(n)}}\hat{b}\ket{n',l'_{(n')}}\bra{n',l'_{(n')}}.
\end{aligned}
\end{equation}
The matrix element of $\hat{b}$ is evaluated via the polaron transformation $U = e^{-S}$ with $S = \hat{N}_c\beta_0(\hat{b}^\dagger - \hat{b})$. Since $[S,\hat{b}] = -\beta_0\hat{N}_c$ and all higher commutators vanish, the Baker--Campbell--Hausdorff series terminates at first order, giving $U\hat{b}\,U^\dagger = \hat{b} + \beta_0\hat{N}_c$. Hence
\begin{equation}\label{c0b}
\begin{aligned}
\bra{n,l_{(n)}}\hat{b}\ket{n',l'_{(n')}} &= \bra{n,l}U\hat{b}\,U^\dagger\ket{n',l'}\\
&= \bra{n,l}(\hat{b} + \beta_0\hat{N}_c)\ket{n',l'}\\
&= \delta_{n,n'}\bigl[\bra{l}\hat{b}\ket{l'} + \beta_0 n\,\delta_{l,l'}\bigr].
\end{aligned}
\end{equation}
The matrix element is diagonal in photon number: $\hat{b}$ does not change $n$. It separates into an off-diagonal part in the mechanical index (the jump operator) and a diagonal part (the static displacement $\beta_0 n$). Substituting Equation~(\ref{c0b}) into Equation~(\ref{c0a}) and using $E_{n,l} - E_{n,l'} = (l-l')\omega_m$,
\begin{equation}\label{c0c}
\begin{aligned}
\hat{b}(t) &= \sum_{n,l,l'} e^{i\omega_m(l-l')t}\; \ket{n,l_{(n)}}\Bigl[\bra{l}\hat{b}\ket{l'} + \beta_0 n\,\delta_{l,l'}\Bigr]\bra{n,l'_{(n)}}\\
&= \underbrace{\sum_{n,l,l'} e^{i\omega_m(l-l')t}\,\ket{n,l_{(n)}}\bra{l}\hat{b}\ket{l'}\bra{n,l'_{(n)}}}_{\text{Term~A}} \;+\; \underbrace{\sum_{n,l}\beta_0 n\,\ket{n,l_{(n)}}\bra{n,l_{(n)}}}_{\text{Term~B}}.
\end{aligned}
\end{equation}
In Term~A, $\bra{l}\hat{b}\ket{l'}$ is non-zero only for $l' = l+1$, so $e^{i\omega_m(l-l')t} = e^{-i\omega_m t}$ for every term. The remaining operator sum is the off-diagonal-in-$l$ part of $\hat{b}$ in the dressed basis. Since the full operator decomposes as $\hat{b}|_{\text{dressed}} = \hat{b}_{\text{off-diag}} + \beta_0\hat{N}_c$, we have $\hat{b}_{\text{off-diag}} = \hat{b} - \beta_0\hat{N}_c$, yielding Term~A $= e^{-i\omega_m t}(\hat{b} - \beta_0\hat{N}_c)$. Term~B is simply $\beta_0\hat{N}_c$. Therefore
\begin{equation}\label{c0d}
\hat{b}(t) = e^{-i\omega_m t}(\hat{b} - \beta_0\hat{N}_c) + \beta_0\hat{N}_c,
\end{equation}
which is Equation~(\ref{6b}). The physical content is that $\hat{b}(t)$ resolves into a component rotating at $\omega_m$ (quantum fluctuations that exchange energy with the bath) and a zero-frequency component $\beta_0\hat{N}_c$ (the static radiation-pressure displacement that shifts energy levels without energy exchange). The DSME uses each component with the correct spectral density: the rotating part couples to $J(\omega_m)$, and the static part couples to $J(0)$, which is the origin of the temperature-dependent dephasing term in Equation~(\ref{5}). The SME, by contrast, applies the same spectral weight to both components because it does not distinguish their distinct frequency content (see Appendix~\ref{C2}).

\section{Mechanical damping}\label{C1}
The cavity-field damping calculations parallel the standard squeezed-vacuum-reservoir treatment for $\hat{a}$ and $\hat{a}^\dagger$ found in standard quantum optics textbooks, and are not reproduced here. We focus on the mechanical operators. From Equation~(\ref{9}) in Section \ref{sec5a},
\begin{equation}\label{c1}
\int_{0}^{\infty} \mathcal{R}_M(s,t)\,e^{i\omega_m s} \,ds = \pi|G(\omega_m)|^2\Big[(N_{eff}+1)-M_{eff}^*\,e^{2i\omega_m t}\Big]
\end{equation}
\begin{equation}\label{c2}
\int_{0}^{\infty} \mathcal{R}_M(s,t)\,e^{-i\omega_m s} \,ds = \pi|G(\omega_m)|^2\Big[N_{eff}-M_{eff}\,e^{-2i\omega_m t}\Big]
\end{equation}
\begin{equation}\label{c3}
\int_{0}^{\infty} \mathcal{R}_M(s,t) \,ds \approx \pi|G(\omega_m)|^2\Big[(N_{eff}+1)-M_{eff}^*\Big],
\end{equation}
where we have assumed approximately equal contributions from the spectral density $J(\omega)$ near $\omega=0$ and $\omega=\omega_m$. If the spectral density $J(\omega=0)\approx 0$, the contribution of Equation~(\ref{c3}) is negligible as compared to Equations~(\ref{c1}) and (\ref{c2}). Recalling, $|G(\pm\omega_m)|^2=\pm\gamma_m/(2\pi)$ (Ohmic reservoir), Equation~(\ref{c3}) for high temperatures ($k_B T\gg\omega_m$) becomes
\begin{equation}\label{c4}
\int_{0}^{\infty} \mathcal{R}_M(s,t) \,ds =\frac{\gamma_m}{2}\left(\frac{k_B T}{\omega_m}\right)\Big[\cosh^2{r}+\sinh^2{r}+2\cosh{r}\sinh{r}\,e^{-i\theta}\Big].
\end{equation}
We define a hermitian operator $\chi_m(t)=\hat{b}(t)+\hat{b}^\dagger(t)$ in the interaction picture. From Equation~(\ref{6b}), $\hat{b}(t) = e^{-i\omega_m t}\hat{B}_m + \beta_0\hat{N}_c$ with $\hat{B}_m = \hat{b} - \beta_0\hat{N}_c$, and $\hat{b}^\dagger(t) = e^{i\omega_m t}\hat{B}_m^\dagger + \beta_0\hat{N}_c$. Cross terms between the rotating and static components carry factors $e^{\pm i\omega_m t}$ and are dropped under the rotating wave approximation used below. Our aim is to calculate terms in
\begin{equation}\label{c5}
\dfrac{d\Tilde{\rho}(t)}{dt}=\int_{0}^{\infty}ds\,\mathcal{R}_M(s,t)\Big[\chi_m(t)\Tilde{\rho}(t)\chi_m(t-s)-\chi_m(t)\,\chi_m(t-s)\,\Tilde{\rho}(t)\Big]+\mathrm{h.c.},
\end{equation}
for a squeezed thermal reservoir. Let us consider the first term in Equation~(\ref{c5}),
\begin{equation}\label{c6}
\begin{split}
    \int_{0}^{\infty}ds\,\mathcal{R}_M(s,t)\,\chi_m(t)\Tilde{\rho}(t)\chi_m(t-s)&=\int_{0}^{\infty}ds\,\mathcal{R}_M(s,t)\Big[\hat{b}(t)\Tilde{\rho}(t)\hat{b}^\dagger(t-s)\\
    &+\hat{b}^\dagger(t)\Tilde{\rho}(t)\hat{b}(t-s)+\hat{b}(t)\Tilde{\rho}(t)\hat{b}(t-s)\\
    &+\hat{b}^\dagger(t)\Tilde{\rho}(t)\hat{b}^\dagger(t-s)\Big].
\end{split}
\end{equation}
Next we use Equations~(\ref{c1}) to (\ref{c4}) to piece-wise calculate the above integral for a squeezed thermal reservoir and neglect fast oscillating terms under the rotating wave approximation (RWA),
\begin{equation}\label{c7}
\begin{split}
    \int_{0}^{\infty}ds\,\mathcal{R}_M(s,t)\,\hat{b}(t)\Tilde{\rho}(t)\hat{b}^\dagger(t-s)&=\frac{\gamma_m}{2}\Big[N_{eff}\hat{B}_m\Tilde{\rho}(t)\hat{B}_m^\dagger\\
    &+\big((N_{eff}+1)-M_{eff}^*\big)\beta_0^2\,\hat{N}_c\Tilde{\rho}(t)\hat{N}_c\Big],
\end{split}
\end{equation}
\begin{equation}\label{c8}
\begin{split}
    \int_{0}^{\infty}ds\,\mathcal{R}_M(s,t)\,\hat{b}^\dagger(t)\Tilde{\rho}(t)\hat{b}(t-s)&=\frac{\gamma_m}{2}\Big[(N_{eff}+1)\hat{B}_m^\dagger\Tilde{\rho}(t)\hat{B}_m\\
    &+\big((N_{eff}+1)-M_{eff}^*\big)\beta_0^2\,\hat{N}_c\Tilde{\rho}(t)\hat{N}_c\Big],
\end{split}
\end{equation}
\begin{equation}\label{c9}
\begin{split}
    \int_{0}^{\infty}ds\,\mathcal{R}_M(s,t)\hat{b}(t)\Tilde{\rho}(t)\hat{b}(t-s)&=\frac{\gamma_m}{2}\Big[M_{eff}^*\hat{B}_m\Tilde{\rho}(t)\hat{B}_m\\
    &+\big((N_{eff}+1)-M_{eff}^*\big)\beta_0^2\,\hat{N}_c\Tilde{\rho}(t)\hat{N}_c\Big],
\end{split}
\end{equation}
\begin{equation}\label{c10}
\begin{split}
    \int_{0}^{\infty}ds\,\mathcal{R}_M(s,t)\hat{b}^\dagger(t)\Tilde{\rho}(t)\hat{b}^\dagger(t-s)&=\frac{\gamma_m}{2}\Big[M_{eff}\hat{B}_m^\dagger\Tilde{\rho}(t)\hat{B}_m^\dagger\\
    &+\big((N_{eff}+1)-M_{eff}^*\big)\beta_0^2\,\hat{N}_c\Tilde{\rho}(t)\hat{N}_c\Big].
\end{split}
\end{equation}
Substituting the above equations in Equation~(\ref{c5}) and similarly calculating the Hermitian conjugate, we get the mechanical part of Equation~(\ref{10}).

\section{Cavity dephasing in SME}\label{C2}
Emergence of cavity dephasing terms in Equation~(\ref{13}) can be understood as follows. The unitary transformation of the dissipator $\mathcal{D}[\hat{b}]\rho(t)$ from Schr\"odinger picture of the standard master equation (SME) to its interaction picture in dressed basis is
\begin{equation}\label{c11}
U^\dagger(t)\mathcal{D}[\hat{b}]\rho(t)U(t)\approx\mathcal{D}[\hat{b}-\beta_0\hat{N}_c]\Tilde{\rho}(t)+\mathcal{D}[\beta_0\hat{N}_c]\Tilde{\rho}(t),
\end{equation}
where $U=\exp{-i\,H't}$, $H'$ being the system Hamiltonian in Equation~(\ref{eq:hamiltonian_transformation}). The expansion of the transformed dissipator also generates cross terms between the rotating component $\hat{B}_m$ and the static component $\beta_0\hat{N}_c$. These oscillate at $\pm\omega_m$ and are dropped in the rotating wave approximation, and the surviving terms are to be read in the Lindblad closure stated after Equation~(\ref{eq:bogoliubov}). Similarly for a squeezed thermal reservoir,
\begin{equation}\label{c12}
\begin{aligned}
U^\dagger(t)\Big(M_{eff}^*\,\hat{b}\rho(t)\hat{b}+M_{eff}\,\hat{b}^\dagger\rho(t)\hat{b}^\dagger\Big)U(t)&\approx M_{eff}^*\hat{B}_m\Tilde{\rho}(t)\hat{B}_m+M_{eff}\hat{B}_m^\dagger\Tilde{\rho}(t)\hat{B}_m^\dagger\\
&+\beta_0^2\Big(M_{eff}^*+M_{eff}\Big)\mathcal{D}[\hat{N}_c]\Tilde{\rho}(t),
\end{aligned}
\end{equation}
where $\hat{B}_m=\hat{b}-\beta_0\hat{N}_c$. These two relations supply exactly the additional terms of Equation~(\ref{13}). The $\mathcal{D}[\hat{N}_c]$ contribution of Equation~(\ref{c11}), weighted by the prefactors $\gamma_m(N_{eff}+1)$ and $\gamma_m N_{eff}$ of the two jump dissipators, yields the $\gamma_m(2N_{eff}+1)\beta_0^2\,\mathcal{D}[\hat{N}_c]$ term. The $\mathcal{D}[\hat{N}_c]$ contribution of Equation~(\ref{c12}), with $M_{eff}^*+M_{eff}=-2\cosh{r}\sinh{r}\cos{\theta}\,(2n_{\text{th}}+1)$ and the overall minus sign of the phase-sensitive terms, yields the $2\gamma_m\cosh{r}\sinh{r}\cos{\theta}\,(2n_{\text{th}}+1)\beta_0^2\,\mathcal{D}[\hat{N}_c]$ term.

\end{document}